\documentclass[12pt]{article}
\usepackage{epsfig}

\def\beq{\begin{equation}}
\def\eeq#1{\label{#1}\end{equation}}
\def\eeqn{\end{equation}}

\def\beqa{\begin{eqnarray}}
\def\eeqa#1{\label{#1}\end{eqnarray}}
\def\eeqan{\end{eqnarray}}

\let\bar=\overbar

\def\Dslash{\not{\hbox{\kern-4pt $D$}}}
\def\dslash{\not{\hbox{\kern-2pt $\del$}}}

\def\msb{{\bar{\ssstyle M \kern -1pt S}}}

\def\Title#1{\begin{center} {\Large {\bf #1} } \end{center}}

\usepackage{xspace} 
\usepackage{lineno}
\usepackage{amsmath}
\usepackage{subfig}
\usepackage{mwe}
\newcommand{\decay}[2]{\ensuremath{#1\!\to #2}\xspace}         % {\Pa}{\Pb \Pc}
\def\to                 {\ensuremath{\rightarrow}\xspace}

\def\Ppi         {\ensuremath{\pi}\xspace}

\def\Pb      {\ensuremath{b}\xspace}                 
\def\Pp      {\ensuremath{p}\xspace}     

\def\pion   {{\ensuremath{\Ppi}}\xspace}

\def\PLambda      {\ensuremath{\Lambda}\xspace}   
\def\Lz          {{\ensuremath{\PLambda}}\xspace}

\def\pip    {{\ensuremath{\pion^+}}\xspace}
\def\pim    {{\ensuremath{\pion^-}}\xspace}

\def\proton      {{\ensuremath{\Pp}}\xspace}

\def\bquark    {{\ensuremath{\Pb}}\xspace}
\def\Lb      {{\ensuremath{\Lz^0_\bquark}}\xspace}

\def\CP                {{\ensuremath{C\!P}}\xspace}
\def\P                 {{\ensuremath{\!P}}\xspace}

\def\Lbppipipi {\decay{\Lb}{\proton\pim\pip\pim}}

\begin{document}

\Title{CP violation measurements in b hadrons at LHCb \\
\vspace{0.4cm}
\footnotesize{Talk presented at the APS Division of Particles and Fields Meeting (DPF 2017),\\ July 31-August 4, 2017, Fermilab. C170731}}

\bigskip\bigskip

%+\addtocontents{toc}{{\it D. Reggiano}}
%+\label{ReggianoStart}

\begin{raggedright}  

{\it Mar\'ia Vieites D\'iaz\index{Vieites Diaz, M..}\\
Department of Particle Physics\\
Universidade de Santiago de Compostela\\
15706 Satiago de Compostela, SPAIN}
\bigskip\bigskip
\end{raggedright}

\section{Introduction}
The very different amounts of matter and antimatter found in the known universe remains to be one of the most important challengers to the Standard Model (SM) expectations and a strong evidence of the interplay between physics processes beyond the SM and Charge-Parity ($\mathcal{CP}$) violation effects. 

The SM absorbs these effects in quark mixing processes, which are described by a 3$\times$3 unitary matrix, named after the physicists Cabibbo, Kobayashi and Maskawa (CKM) Ref. \cite{ckm,ckm2}. Being a fourth SM-like family excluded by the physics results on the nature of the Higgs and Z bosons from the LHC and LEP data Ref. \cite{fourthfamily}, respectively, the CKM matrix must be unitary in order to conserve probability when quarks mix flavours. The values of the matrix elements are related to different quark pair couplings and can not be predicted by the theory. 

Using the standard parametrisation of the CKM matrix elements, its four degrees of freedom can be identified with three real quantities (rotation angles) and a non zero phase that allows for $\mathcal{CP}$ violation in the model. All these four parameters are over-constrained in the SM framework, providing an excellent scenario to search for incompatibilities and precision tests of the SM implications.

$\mathcal{CP}$ violation effects may occur in nature via three different types of processes. They are classified depending on the nature of two interfering amplitudes, which must have differing strong and weak phases, as mixing, decay, or interference between both. \CP violation manifests in the mixing if the rate at which a neutral meson oscillates to its antiparticle is different to the contrary process, $\mathcal{P}(X\rightarrow \overline{X})\neq\mathcal{P}(\overline{X}\rightarrow X)$. If the two interfering amplitudes in a decay represent two different paths to the same final state, non conserving \CP effects would imply that a given process and its \CP conjugate occur at different rates: $\mathcal{A}(X\rightarrow f) \neq \mathcal{A}(\overline{X} \rightarrow \overline{f})$. The third phenomenon involves interference between the direct decay path and the decay after mixing has occurred. In these cases, partial decay widths are sensitive to the phase difference between the mixing and the decay and a decay-time dependent \CP asymmetry measurement must be performed to disentangle the inferring amplitudes. 

The analyses presented here have used the full data sample from Run 1 of LHCb, corresponding to 1$fb^{-1}$ at $\sqrt{s} = 7$ TeV and 2$fb^{-1}$ at $\sqrt{s} = 8$ TeV of $pp$ collisions. One of the updates includes also 2$fb^{-1}$ at $\sqrt{s} = 13$ TeV from the first part of the LHCb Run 2. The LHCb detector consists of a single-arm forward spectrometer specifically designed for flavour physics. The characteristics of this design include very good vertex and tracking resolution, a fully instrumented forward coverage that maximizes the acceptance of the $b\overline{b}$ quark pairs produced at the LHC and a very efficient particle identification (PID) system separating protons, pions and kaons in the full acceptance. Together with a high performing trigger, reconstruction and PID efficiencies are crucial in order to study flavour physics in the very complicated hadronic environment of the LHC. Further details about the detector can be found in Ref. \cite{lhcb}.

\section{Status of $2\beta$ and $2\beta_s$ measurements}
These two angles are accessible from the interference between mixing and decay, the subindex indicating whether the initial decaying meson was a $B^0$ or a $B^0_s$. The SM predicts $\phi^{c\overline{c}s}_s \equiv -2\beta_s \equiv -2 \rm{arg}\left(-\frac{V_{ts}V^*_{tb}}{V_{cs}V^*_{cb}}\right) = 0.0365^{+0.0013}_{-0.0012}$ rad and $\phi^{c\overline{c}s}_d \equiv 2\beta \equiv -2 \rm{arg}\left(-\frac{V_{cd}V^*_{cb}}{V_{td}V^*_{tb}}\right) = 0.771^{+0.017}_{-0.041}$ rad, for the $B^0$ and the $B^0_s$ systems, respectively (Ref. \cite{CKMfitter}). These phases are accessible experimentally measuring the time-dependent \CP violating asymmetries of the so called golden modes, the $B^0_s\rightarrow J/\psi K^+K^-$ and the $B^0\rightarrow J/\psi K^0_s$ decay channels. Table~\ref{tab:a1} summarises the latest results by the LHCb collaboration on these two phases. The following sections describe two of the latest published analyses in more detail. 

\begin{table}[b]
\begin{footnotesize}
\begin{center}
\begin{tabular}{l| c r}	

	Decay & Result & Reference\\
	\hline
	$B^0_s\rightarrow J/\psi \pi^+\pi^-$ & +0.070$\pm$0.068$\pm$0.008&PLB B736 186 (2014)\\
	$B^0_s \rightarrow D^+_sD^-_s$ &+0.02$\pm$0.17$\pm$0.02 &PRL113 211801 (2014)\\
	$B^0\rightarrow J/\psi \pi^+\pi^-$ & $\Delta2\beta_f = (-0.9\pm9.7^{+2.8}_{-6.3})^\circ$ & PLB 742 (2015)\\
	$B^0_s\rightarrow J/\psi K^+K^-$ &-0.058$\pm$0.049$\pm$0.006 &PRL114 041802 (2015)\\
	$B^0 \rightarrow J/\psi K_S^0$ & $\sin(2\beta) = 0.746\pm 0.030 $ & PRL 115, 031601 (2015)\\
	$B^0 \rightarrow D^+D^-$ & $\Delta\phi=-0.16^{+0.19}_{-0.21}$ & PRL 117 261801(2016)\\
	$B^0_s \rightarrow \psi(2S)\phi$ &$+0.23^{+0.29}_{-0.28}\pm0.02$ &PLB B762, 252-262 (2016)\\
	$B^0_s\rightarrow J/\psi K^+K^-$ ($m_{K^+K^-}>m_{\phi(1020)}$) &+0.119 $\pm$0.107$\pm$0.034 & JHEP 08 (2017) 037\\
	$B^0\rightarrow J/\psi K^0_S$ & $\sin(2\beta) = 0.83 \pm 0.08 \pm 0.01 $ & JHEP (LHCB-PAPER-2017-029) \\
	$B^0\rightarrow \psi(2S) K^0_S$ & $\sin(2\beta) = 0.84 \pm 0.10 \pm 0.01 $ & JHEP (LHCB-PAPER-2017-029) \\
		\hline

\end{tabular}
\caption{Latest published results on $\phi_{(s,d)}$ by the LHCb collaboration. \textit{Note:} results on $\sin(2\beta)$ from the $B^0\rightarrow J/\psi (\rightarrow e^+e^-) K^0_S$ and $B^0\rightarrow \psi(2S)(\rightarrow \mu^+\mu^-) K^0_S$ decay channels are included for completeness, but they are not discussed in the text as they were not public at the time of the conference.}
\label{tab:a1}
\end{center}
\end{footnotesize}
\end{table}

\subsection{Measurement of $\phi^{c\overline{c}s}_s$ in $B^0_s\rightarrow J/\Psi K^+K^-$}
This is a flavour-tagged, time-dependent amplitude analysis, in which the invariant mass of the two kaons system is restricted to the region above the $\phi(1020)$ resonance. The high invariant mass spectrum of the two kaons system presents a rich structure in terms of high spin resonances, being this the first study performed with final state data dominated by a tensor. 

In order to be considered, events must fulfil loose kinematic requirements and PID criteria. Random combination of tracks remaining in the selected sample are suppressed by a multivariate classifier. At this stage, the data sample is dominated by signal candidates and per event weights are assigned using the $s$Fit Ref. \cite{sfit} technique in order to subtract background and perform the analysis with a signal-only model. 

\begin{figure}[htb]
\begin{center}
\includegraphics[width=0.85\textwidth]{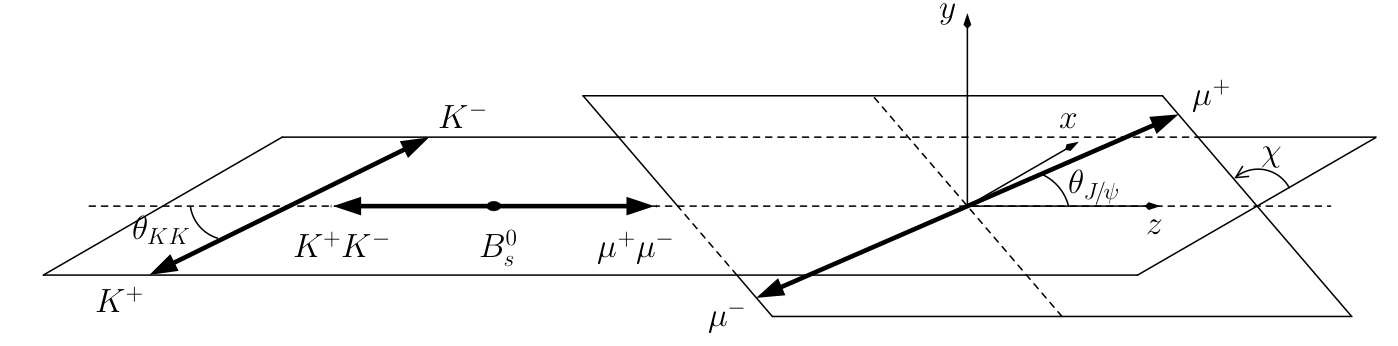}
\caption{Helicity angles definition followed in the $B^0_s\rightarrow J/\Psi K^+K^-$ analysis.}
\label{fig:angles}
\end{center}
\end{figure}

The final fit is performed in a five dimensional space defined by the decay time, the invariant mass of the two kaons system and three helicity angles. These angles are shown schematically in Figure~\ref{fig:angles} and consist of the angle $\theta_{KK}$ between the $K^+$ direction in the $K^+K^-$ rest frame with respect to the $K^+K^-$ direction in the $B^0_s$ rest frame, the angle $\theta_{J/\psi}$ between the $\mu^+$ direction in the $J/\psi$ rest frame with respect to the $J/\psi$ direction in the $B^0_s$ rest frame, and the angle $\chi$ between the $J/\psi$ and $K^+K^-$ decay planes in the $B^0_s$ rest frame. 

The reconstruction and selection efficiencies as a function of the decay time are measured in the $B^0\rightarrow J/\psi K^{*0}(\rightarrow K^+\pi^-)$ control channel. The small kinematic differences between signal candidates and the control mode are taken into account in the efficiency computation. The LHCb detector acceptance, as well as the momentum requirements on the final state particles influence the distributions of the fitted variables. These effects are parameterised relying on simulated data using combinations of Legendre polynomials ($\theta_{KK}$) and spherical harmonics ($\theta_{J/\psi}, \chi$).

\begin{figure}[htb]
\begin{center}
\includegraphics[width=0.85\textwidth]{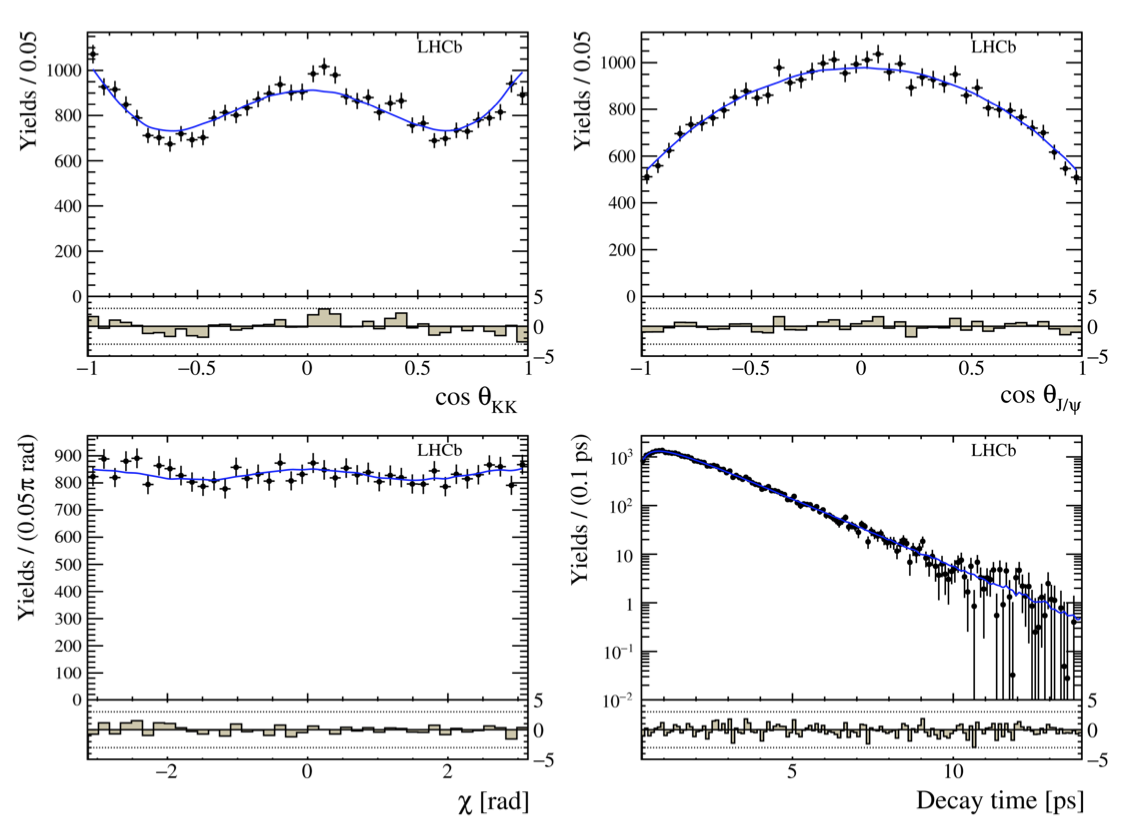}
\caption{Fit projections superimposed with data distributions for the three helicity angles and the decay time in the $B^0_s\rightarrow J/\Psi K^+K^-$.}
\label{fig:fitproj}
\end{center}
\end{figure}

\begin{figure}[htb]
\begin{center}
\includegraphics[width=0.65\textwidth]{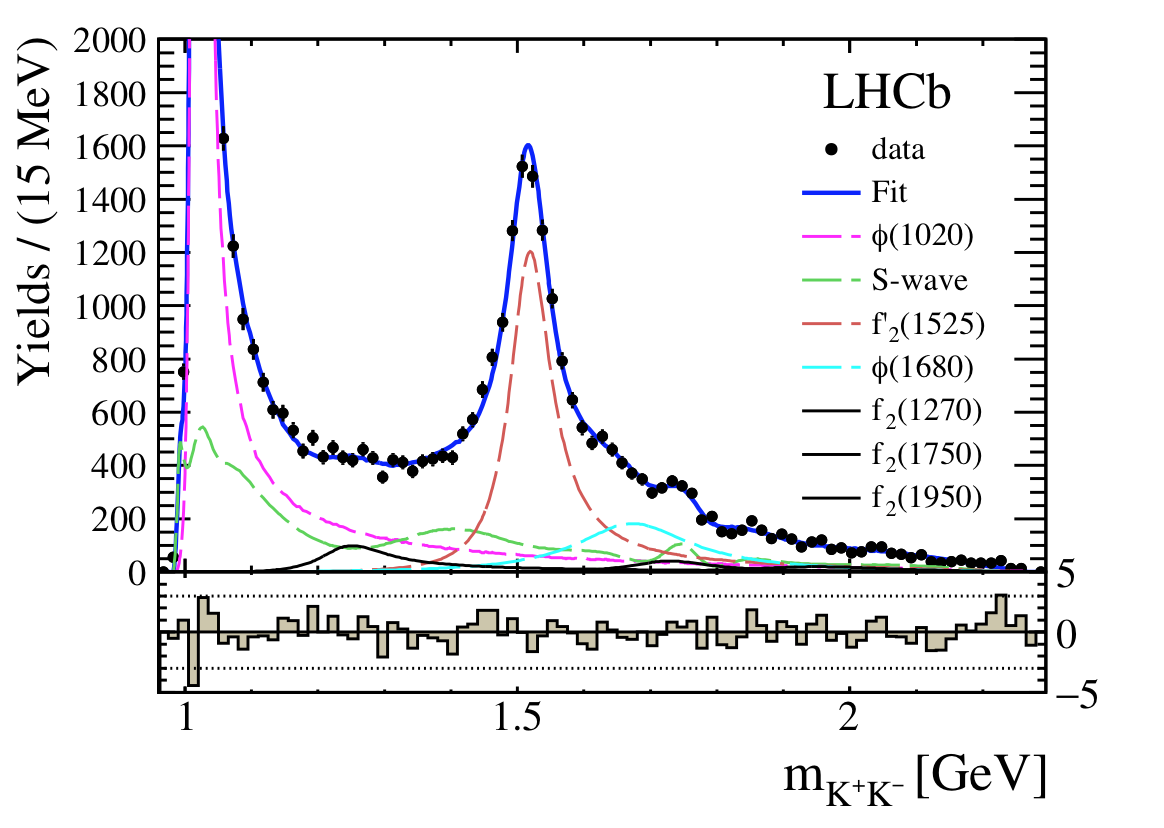}
\caption{Projection of the invariant mass of the kaon pair in the high (above the $\phi(1020)$ meson) mass region. The different amplitudes accounted for in the model are also plotted.}
\label{fig:2kmass}
\end{center}
\end{figure}

Projections of the decay time and the helicity angles for the high mass (above the $\phi(1020)$ meson) are shown in Figure~\ref{fig:fitproj} and the fit results for $\mathcal{CP}$ violating parameters are summarised in Table~\ref{tab:results1}. The invariant mass spectrum of the two kaons system is shown in Figure~\ref{fig:2kmass}, where the peak from the $f'_2(1525)$ resonance is clearly visible. 

\begin{table}[t]
\begin{footnotesize}
\begin{center}
\begin{tabular}{c| c c c }	
Parameter &  $m_{K^+K^-}>$1.05GeV result & Combination 1 & Combination 2\\
\hline
$\phi_s$ (mrad) & $119 \pm 107 \pm 34$ & $-25 \pm 45 \pm 8$  & 1 $\pm$ 37 \\
$|\lambda|$ & $ 0.994 \pm 0.018 \pm 0.006$& $ 0.978 \pm 0.013 \pm 0.003$ & 0.973 $\pm$ 0.013\\
$\Gamma_s $ (ps$^{-1}$) & $0.650 \pm 0.006 \pm 0.004$  & $0.6588 \pm 0.0022 \pm 0.0015$ & $0.6588 \pm 0.0022 \pm 0.0015$\\
$\Delta\Gamma_s$ (ps$^{-1}$) & $0.066 \pm 0.018 \pm 0.010$ & $0.0813 \pm 0.0073 \pm 0.0036$ & $0.0813 \pm 0.0073 \pm 0.0036$\\
\end{tabular}
\caption{Results of the $B_s^0\rightarrow J/\psi K^+K^-$ analysis: as obtained from the nominal fit in the two kaon invariant mass region $m_{K^+K^-}>$1.05GeV, after making the combination with previous results obtained in the $\phi(1020)$ region Ref. \cite{phiresults} (Combination 1) and after adding more results from $B^0_{(s)}\rightarrow J/\psi \pi^+\pi^-$ decays Ref. \cite{jpsipipi} (Combination 2).}
\label{tab:results1}
\end{center}
\end{footnotesize}
\end{table}

%%%%%%%

\subsection{Time dependent $\mathcal{CP}$ violation in the $B^0_{(s)}\rightarrow h^+h^-$ system}
It has been proven that a combined analysis of the branching ratios and \CP asymmetries in the $B^0_{(s)}\rightarrow h^+h^-$ system, if accounting for U-spin symmetry breaking effects, allows for the determination of the CKM phases $\gamma$ and $-2\beta_s$ Ref. \cite{uspin1,uspin2}. In the present analysis, the study of time-dependent $\mathcal{CP}$-violating asymmetries is performed. Assuming $CPT$ invariance, the $\mathcal{CP}$ asymmetry as a function of time between $B_{(s)}^0$ and $\overline{B}^0_{(s)}$ mesons decaying to a $\mathcal{CP}$ eigenstate $f$ is given by:

\begin{equation}
\mathcal{A}(t) = \frac{\Gamma_{\overline{B}^0_{(s)}\rightarrow f}(t) -  \Gamma_{B^0_{(s)}\rightarrow f}(t)}{\Gamma_{\overline{B}^0_{(s)}\rightarrow f}(t) +  \Gamma_{B^0_{(s)}\rightarrow f}(t) } = \frac{-C_f\cos(\Delta m_{d,s}t) + S_f\sin(\Delta m_{d,s}t)}{\cosh \left(\frac{\Delta\Gamma_{d,s}}{2}t\right) + A^{\Delta\Gamma}_f \sinh\left(\frac{\Delta\Gamma_{d,s}}{2}t\right)}
\end{equation}

where $\Delta m_{d,s}$ and $\Delta\Gamma_{d,∫s}$ are the mass and width differences of the mass eigenstates in the $B^0_{(s)} - \overline{B}^0_{(s)}$ system. The quantities $C_f, S_f$ and $A^{\Delta\Gamma}_f$ are
\begin{equation}
C_f \equiv \frac{1-|\lambda_f|^2}{1+|\lambda_f|^2}, S_f\equiv \frac{2\rm{Im}\lambda_f}{1+|\lambda_f|^2}, A^{\Delta\Gamma}_f \equiv \frac{2\rm{Re}\lambda_f}{1+|\lambda_f|^2},
\end{equation}
where $\lambda_f$ has the usual definition
\begin{equation}
\lambda_f \equiv \frac{q}{p}\frac{\overline{A}_f}{A_f}.
\end{equation}

In the absence of sizeable $\mathcal{CP}$ violation effects arising in the $B^0_{(s)}$ meson mixing (equivalent to assume $|q/p|=1$), the $C_f$ and $S_f$ coefficients parametrise the $\mathcal{CP}$ violation in the decay and in the interference between mixing and decay, respectively. Furthermore, when added in quadrature and considering $A^{\Delta\Gamma}_f$ as well, these parameters satisfy the relation $|C_f|^2 + |S_f|^2 +|A^{\Delta\Gamma}_f|^2 = 1$. This constraint was not applied in the present analysis to allow for future combination with results from other analyses. 

Candidates selection begins with hardware and software trigger requirements. The former implies having large transverse energy clusters in the hadronic calorimeter, and the latter a well reconstructed and displaced secondary vertex and, at least, one charged track with a large impact parameter with respect to all the $pp$ interaction points above given thresholds. A multivariate classifier is used offline in order to compute for each candidate its likelihood for being signal or background. The PID hypothesis is used to classify the events in orthogonal subsamples, according to the final state particles  ($K\pi$,$\pi\pi$ and $KK$). The PID calibration is performed following a data-driven method, which uses samples of pions and kaons from the easily identified $D^{*+}$ decays. At the final step of the selection process, the classifier requirement is optimised to maximise the ratio $S/\sqrt{S+B}$ in a $\pm$ 60 MeV window around the nominal mass of the $B^0_{(s)}$, where $S$ stands for the number of signal events and $B$ for the background yield.

The $K\pi$ sample, originated from decays of either $B^0$ or $B^0_s$ mesons, is used as calibration to determine the flavour of the decaying $B^0_{(s)}$ mesons at production time. At this time, only one type of the LHCb flavour taggers, the so-called opposite sign tagger, is being used to determine the flavour of the $B^0_{(s)}$ mesons at production time. Important updates are therefore expected once all the available taggers are included in the analysis strategy.

\begin{figure}[htb]
\begin{center}
\includegraphics[width=0.45\textwidth]{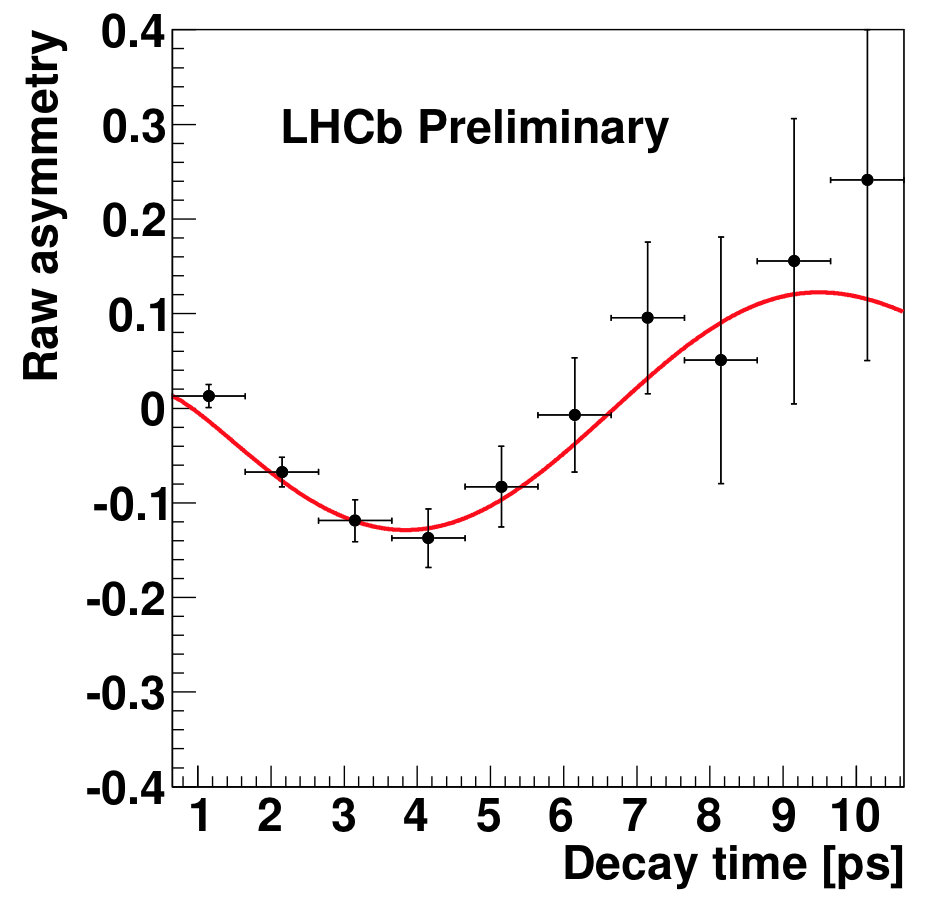}
\includegraphics[width=0.45\textwidth]{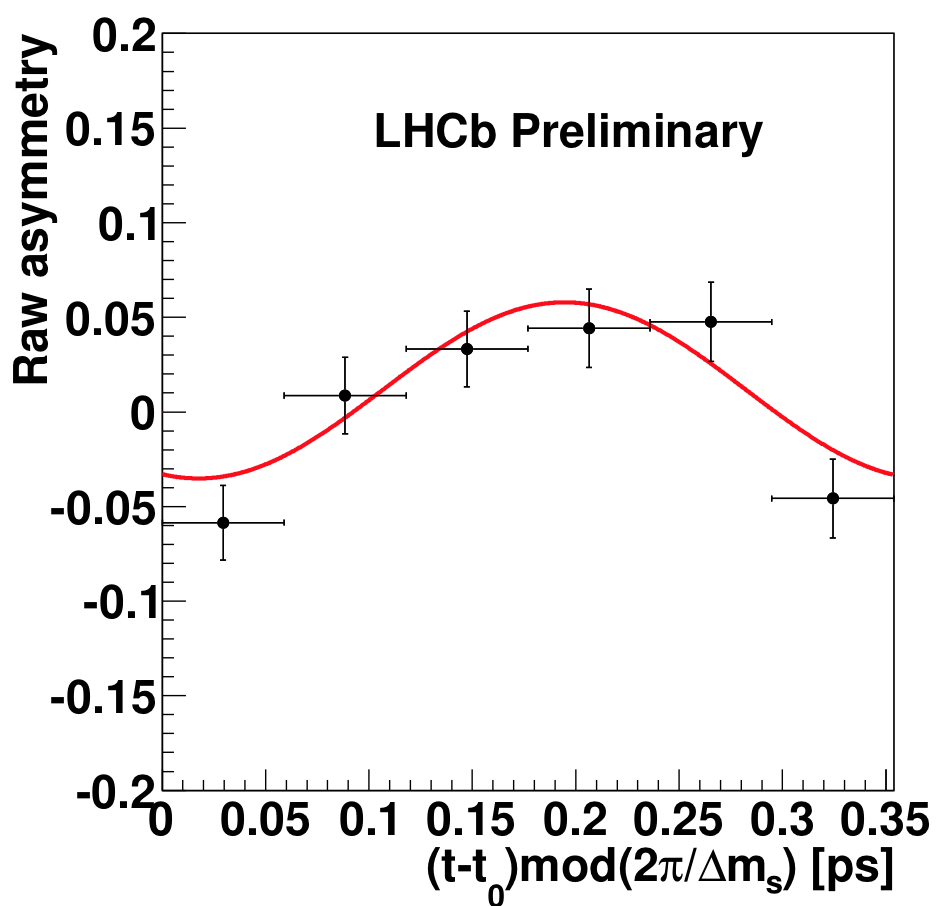}
\caption{Time-dependent asymmetry for the $\pi^+\pi^-$ spectrum (left), for candidates lying in the mass region 5.20 $<$ m $<$ 5.35 GeV/c$^2$, and for the $K^+K^-$ spectrum (right) in 5.30 $<$ m $<$ 5.45 GeV/c$^2$.}
\label{fig:bhh}
\end{center}
\end{figure}

A simultaneous fit to the invariant mass, decay time, per-event mistag probability and per-event decay time error distributions of the three subsamples allows to determine $C_{\pi^+\pi^-}, S_{\pi^+\pi^-}, C_{K^+K^-}, S_{K^+K^-}$ and $A_{K^+K^-}^{\Delta\Gamma}$. Three sources of background are modelled in addition to the signal shapes: combinatorial (random combination of tracks), cross feed from misidentified decays and partially reconstructed three-body decays. Several sources of systematic uncertainties are considered, being the dominant ones related to the shapes of the two-body invariant mass distributions and the time resolution calibration parameters. The uncertainties of final results are statistically dominated, except for the $A_{K^+K^-}^{\Delta\Gamma}$ case. The measured values for these observables are shown in Table~\ref{tab:bhh} and the time-dependent asymmetries for the $\pi^+\pi^-$ and $K^+K^-$ yields can be seen in Figure~\ref{fig:bhh}. 

\begin{table}[b]
\begin{footnotesize}
\begin{center}
\begin{tabular}{c| c c c }	
Parameter &  Value\\
\hline
$C_{\pi^+\pi^-}$ & -0.24 $\pm$ 0.07 $\pm$ 0.01\\
$S_{\pi^+\pi^-}$& -0.68 $\pm$ 0.06 $\pm$ 0.01\\
$C_{K^+K^-}$& 0.24 $\pm$ 0.06 $\pm$ 0.02\\
$S_{K^+K^-}$ &0.22 $\pm$ 0.06 $\pm$ 0.02\\
$A_{K^+K^-}^{\Delta\Gamma}$&-0.75 $\pm$ 0.07 $\pm$ 0.11\\
\end{tabular}
\caption{Measured values of the $\mathcal{CP}$ violating parameters in the $B^0_{(s)}\rightarrow h^+h^-$ system time-dependent decay amplitude. }
\label{tab:bhh}
\end{center}
\end{footnotesize}
\end{table}

\section{Status of $\gamma$ measurements}
The $\gamma$ angle is defined as $\gamma = arg\left(-\frac{V_{ud}V^*_{ub}}{V_{cd}V^*_{cb}}\right)$, where the absence of couplings with the top quark should be noticed. This implies that this angle can be directly determined at tree level, where is safe to assume (provided that New Physics does not contribute at tree-level, Ref. \cite{noNPtree}) that there are no sizeable contributions from  non-SM $\mathcal{CP}$ violating processes. An indirect extraction of its value is also possible from global CKM fitters, which do account for higher order diagrams as well. At present time, the value of $\gamma$ obtained from direct determination yields $\gamma = (72.1^{+5.4}_{-5.8})^\circ$, while the indirect constraints indicate $\gamma = (65.3^{+1.0}_{-2.5})^\circ$ Ref. \cite{CKMfitter}. Further constraining both results and elucidating the tension between them will give great insight into the SM consistency checks. 

\begin{figure}[htb]
\begin{center}
\includegraphics[width=0.85\textwidth]{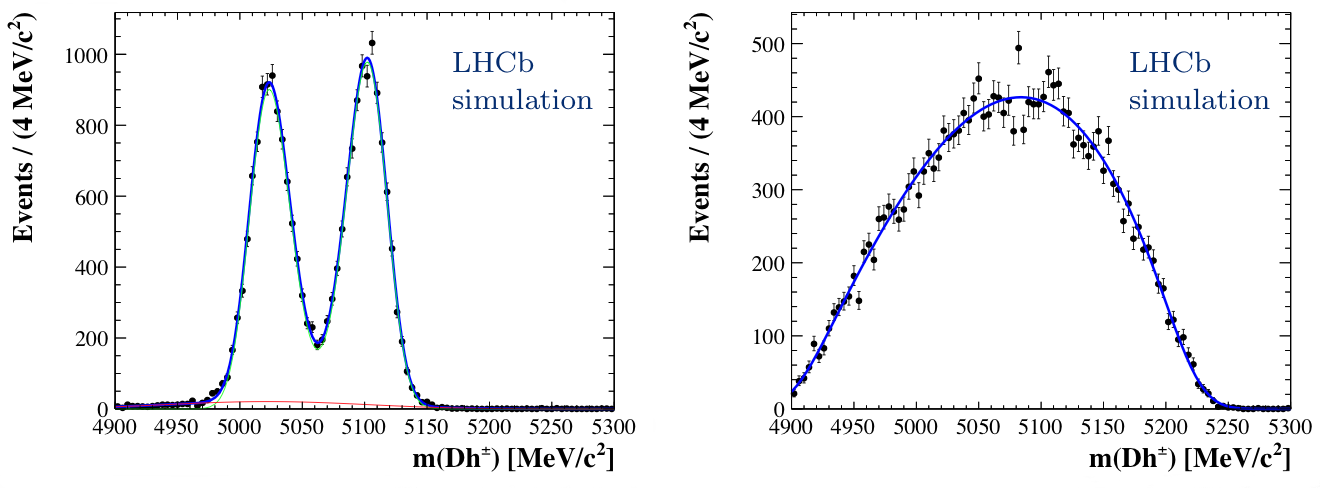}
\caption{Distributions of $m(D^0h^\pm)$ obtained from simulated data for the $B^\pm\rightarrow D^{*0}(\rightarrow D^0\pi^0)h^\pm$ (left) and the $B^\pm\rightarrow D^{*0}(\rightarrow D^0\gamma)h^\pm$ (right) decay modes.}
\label{fig:gammapi0}
\end{center}
\end{figure}

The LHCb collaboration has recently performed an updated combination of several $\gamma$ measurements from different decay channels. The same procedure that was followed to obtain the previous result Ref. \cite{prevgamma} was used in this new combination. Among the new inputs, the particular decay topology of $B^\pm \rightarrow D^{*0}h^\pm$ decays was explored for the first time. From the theoretical point of view, this mode is very similar to the well understood $B^\pm \rightarrow D^{0}h^\pm$ channel. On the other hand, from the experimental side, and in particular considering the LHCb detector, the reconstruction of the $D^{*0}$ meson presents an important drawback due to the presence of either a neutral pion or a photon among the final state particles. This potential challenge is solved by profiting from the different observed in the kinematics of the $D^0$ meson itself depending on which particle it was produced in association with. In the $D^{*0}\rightarrow D^0\gamma$ case, the spin of the photon in the final state implies that the decay products must be totally polarised in order to conserve total angular momentum, while for the pseudoscalar $\pi^0$ this restriction does not appear in the $D^{*0}\rightarrow D^0\pi^0$ decay. This difference gives raise to two clearly differentiated $m(D^0h^\pm)$ invariant mass distributions, as shown in Figure~\ref{fig:gammapi0}. This also implies that there is no need to reconstruct all the decay products from the $D^{*0}$ meson, which also allows to use the same selection requirements that are used for the $B^\pm \rightarrow D^{0}h^\pm$ analyses.

\begin{figure}[htb]
\begin{center}
\includegraphics[width=0.85\textwidth]{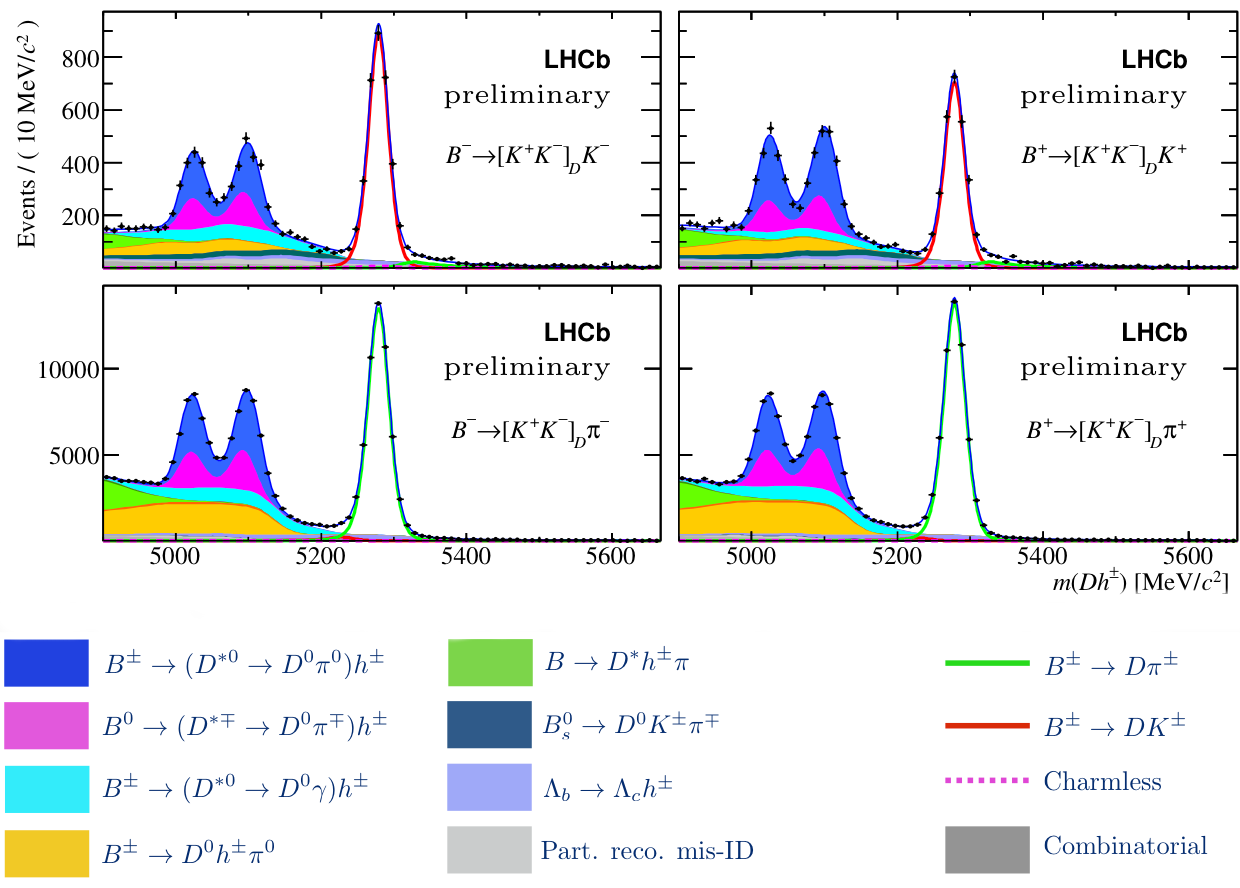}
\caption{Invariant mass distributions of selected $B^\pm\rightarrow [K^-K^+]_{D^0}h^\pm$ candidates, split by the charge of the decaying $B^\pm$ meson and the nature of the bachelor kaon.}
\label{fig:Dkpifit}
\end{center}
\end{figure}

A combined analysis of both the $B^\pm \rightarrow D^{0}h^\pm$ (fully reconstructed) and the $B^\pm \rightarrow D^{*0}h^\pm$ (partially reconstructed) is therefore performed, allowing for an updated result with the full LHCb Run 1 data sample for the fully reconstructed mode. The final fit is performed simultaneously on 12 disjoint samples (data are split depending on the charge of the decaying $B^\pm$ meson, the accompanying hadron $h^\pm\equiv K^\pm, \pi^\pm$ and the three different final states $[K^-\pi^+]_{D^0}, [K^+K^-]_{D^0}, [\pi^+\pi^-]_{D^0}$ considered for the $D^0$ meson), and accesses nineteen $\mathcal{CP}$ violating observables. The fit projections for the $[K^-K^+]_{D^0}$ subsample can be seen in Figure~\ref{fig:Dkpifit}. Further details on the construction of the observables and insight of the analysis can be consulted in Ref. \cite{analysis1}.

\begin{figure}[htb]
\begin{center}
\includegraphics[width=0.65\textwidth]{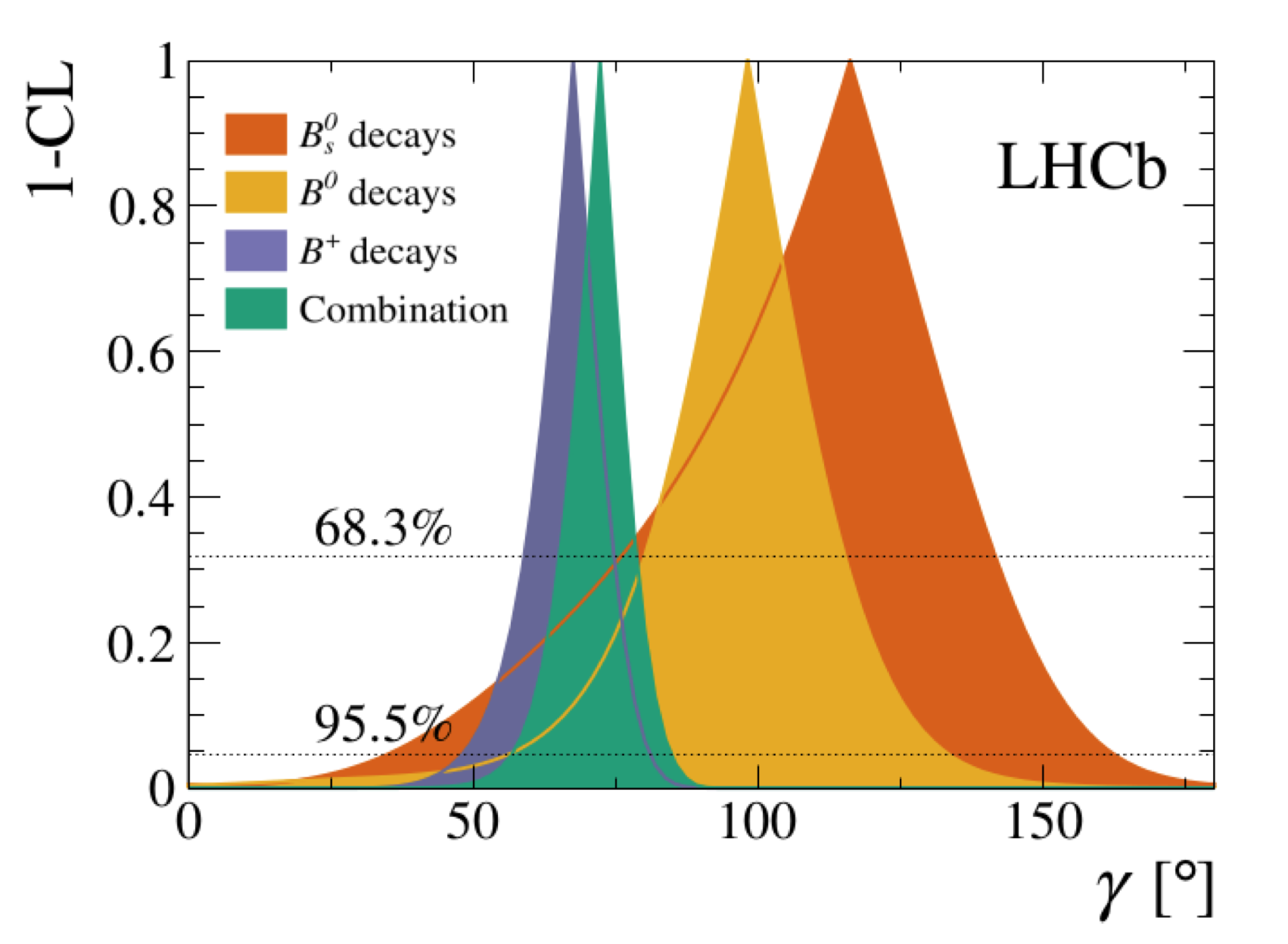}
\caption{1-CL (confidence level) plots, using the profile likelihood method, for the combination of $\gamma$ measurements split by the decaying B meson flavour.}
\label{fig:gamma}
\end{center}
\end{figure}

Including these and other updates ($B^0_s\rightarrow D^\mp_sK^\pm$ analysed with full Run 1 sample Ref. \cite{conf015} and results using the $B^\pm\rightarrow DK^{*\pm}$ for the first time Ref. \cite{conf014}),  the new constraint on the $\gamma$ value derived by the LHCb collaboration yields $\gamma = (76.8^{+5.1}_{-5.7})^\circ$, where the uncertainty includes statistical and systematic contributions. The impact on the global sensitivity of the different decay channels included in the combination is shown in Figure~\ref{fig:gamma}. This combination leads to a significantly smaller uncertainty than the previous combination and replaces it as the most precise determination of $\gamma$ from a single experiment to date.

\section{$\mathcal{CP}$ violation effects in baryons}
According to the SM predictions, a sizeable $\mathcal{CP}$ violation effect should occur in charmless decays of baryonic matter. Asymmetries as large as $\mathcal{A}_{CP}\sim20\%$ have been predicted for $\Lambda_b^0$ decays Ref. \cite{Lbcpv} and their measurement can be used to test the SM consistency and to place constraints on possible extensions of this model. Evidence of such effects has been seen after performing a search for \P and \CP violation in \Lbppipipi decays using triple product asymmetries Ref. \cite{lbdecays}. 

The observables used for this analysis are constructed as follows:
\begin{enumerate}
\item The triple products in the \Lb rest frame are: $C_{\hat{T}} = \vec{p}_p \cdot (\vec{p}_{h^-} \times \vec{p}_{h^+})\propto \sin \Phi$ and $\overline{C}_{\hat{T}} = \vec{p}_{\overline{p}} \cdot (\vec{p}_{h^+} \times \vec{p}_{h^-})\propto \sin \overline{\Phi}$. The pion with the highest momentum in the \Lb rest frame ($\pi_{\text{fast}}$) is chosen for the computation (see Figure \ref{fig:anglesLb}).

\begin{figure}[htb]
\begin{center}
\includegraphics[width=0.65\textwidth]{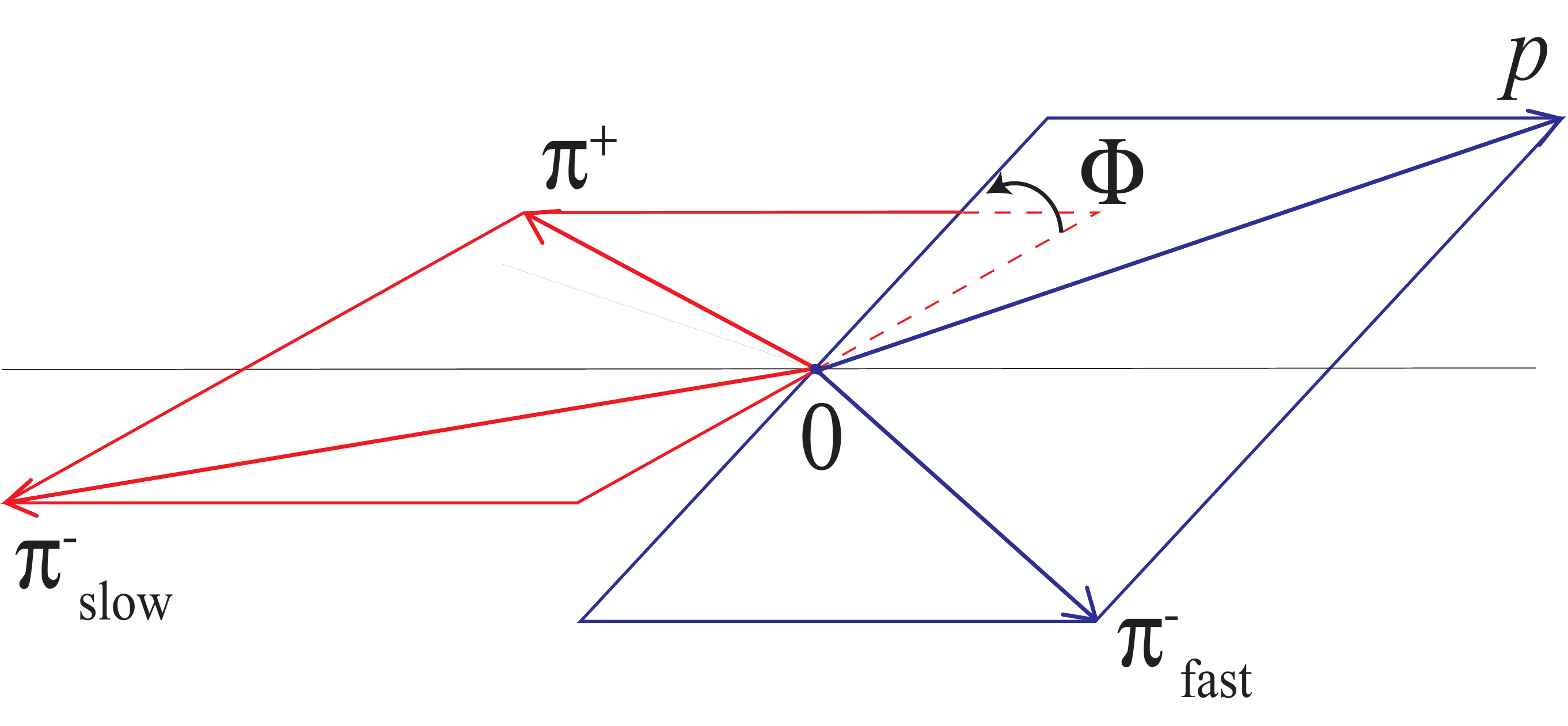}
\caption{Definition of the helicity angles in the \Lb decays. The pion with the highest (lowest) momentum in this reference frame is referred to as $\pi_{\text{fast}}$ ($\pi_{\text{slow}}$).}
\label{fig:anglesLb}
\end{center}
\end{figure}

\item From these, two $\hat{T}-odd$ asymmetries can be built as: 
\begin{equation*}
A_{\hat{T}} = \frac{N_{\Lambda_b^0}(C_{\hat{T}}>0) - N_{\Lambda_b^0}(C_{\hat{T}}<0)}{N_{\Lambda_b^0}(C_{\hat{T}}>0) + N_{\Lambda_b^0}(C_{\hat{T}}<0)}  
  \text{ and } 
\overline{A}_{\hat{T}} = \frac{N_{\overline{\Lambda}_b^0}(-\overline{C}_{\hat{T}}>0) - N_{\overline{\Lambda}_b^0}(-\overline{C}_{\hat{T}}<0)}{N_{\overline{\Lambda}_b^0}(-\overline{C}_{\hat{T}}>0) + N_{\overline{\Lambda}_b^0}(-\overline{C}_{\hat{T}}<0)},
\end{equation*}
where $\hat{T}$ is the motion reversal operator that reverts both momentum and spin three-vectors and $N_{\Lambda_b^0}$ ($N_{\overline{\Lambda}_b^0}$) is the number of $\Lambda_b^0$ ($\overline{\Lambda}_b^0$) signal events. It should be noted that these quantities are, by construction, insensitive to particle/antiparticle production asymmetries and to detector-induced charge asymmetries Ref. \cite{observables}.
\item Finally, the \CP and \P violating observables can be constructed from these asymmetries as:
\begin{equation*}
a^{\hat{T}-odd}_{CP} = \frac{1}{2}(A_{\hat{T}} - \overline{A}_{\hat{T}})
 \qquad \text{and}  \qquad
a^{\hat{T}-odd}_{P} = \frac{1}{2}(A_{\hat{T}} + \overline{A}_{\hat{T}}),
\end{equation*}
where observing a significant deviation from zero in any of their values would indicate a sign for \CP or \P violation, respectively.
\end{enumerate}

\begin{figure}[htb]
\begin{center}
\includegraphics[width=0.85\textwidth]{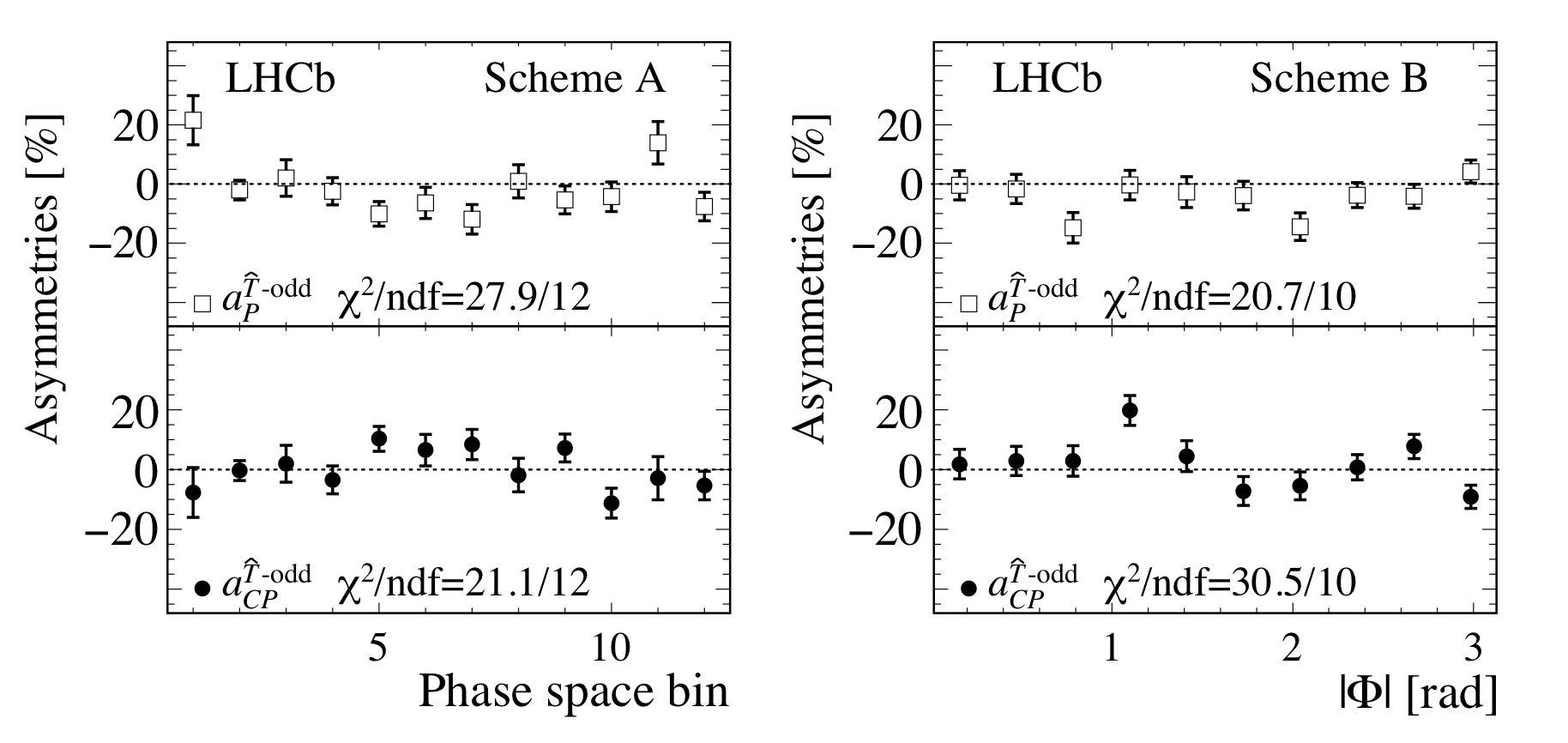}
\caption{Measured values for the \P (upper graph) and \CP (lower graph) violating observables in the \Lbppipipi decay for the two considered binning schemes: phase-space (left) and $|\phi|$ bins (right).}
\label{fig:acp}
\end{center}
\end{figure}

The \P and \CP violating observables are measured in two different binning schemes: one divides the data sample in 10 uniform bins in $\phi$, the other uses known resonances to define the boundaries of 12 non-uniform bins over the phase-space. Figure~\ref{fig:acp} shows the measurements of the \P and \CP asymmetries in these bins. Combining the results of both schemes gives first evidence for \CP violation in \Lbppipipi at the level of 3.3$\sigma$.

\section{Summary and conclusions}
SM predictions remain unbeaten as for what \CP violation results are concerned. Nevertheless, theoretical and experimental uncertainties have both room for improvement. While the first are to be driven by the theoretical community, data taken during LHCb Run 2 and refined analysis techniques will definitely help improving the experimental uncertainties. 

In the present document a selected sample of the latest results on \CP violation by the LHCb collaboration have been presented. These include the first measurement of $\phi_s^{c\overline{c}s} = 0.119 \pm0.107\pm0.034 $ rad in a final state dominated by a tensor resonance, the most precise measurements of $C_{\pi^+\pi^-}$ and $S_{\pi^+\pi^-}$ to date, the first values ever measured of $C_{K^+K^-}$, $S_{K^+K^-}$ and $A_{K^+K^-}^{\Delta\Gamma}$, the latest constraints on $\gamma = (76.8^{+5.1}_{-5.7})^\circ$ from the combination of several LHCb results and the first evidence for \CP violation in baryonic mater at the level of 3.3$\sigma$. All results show a good agreement with the SM expectations.

\end{document}